\documentclass[a4paper,11pt]{article}
\pdfoutput=1 

\usepackage{jinstpub} 

\usepackage{lineno}

\proceeding{12th International Conference on Position Sensitive Detectors - PSD12\\
  12-17 September, 2021\\
  Birmingham, U.K.}

\title{\boldmath ALICE ITS~3: the first truly cylindrical inner tracker}

\author{Domenico Colella}
\affiliation{Politecnico di Bari and INFN sezione di Bari\\ Via E. Orabona 4, Bari, 70125, Italy}
\emailAdd{domenico.colella@ba.infn.it}
\collaboration[c]{on behalf of ALICE Collaboration}

\abstract{
The high integration density of MAPS, with silicon sensor and readout electronics implemented in the same device, allows very thin structures with a greatly reduced material budget. Thicknesses of $\mathcal{O}$(50~$\mu$m), values at which silicon chips become flexible, are readily used in many applications. In addition, MAPS can be produced in sensors of wafer size by a process known as stitching. This in turn allows to build detector elements that are large enough to cover full tracker half-layers with single bent sensors.
The ALICE ITS~3 project is planning to build a new vertex tracker based on truly cylindrical wafer-scale sensors, with <0.05\% X/X$_{0}$ per layer and located as close as 18 mm to the interaction point. R\&D on all project aspects (including mechanics for bent wafer-scale devices, test beams of bent MAPS, design of stitched sensors) is rapidly progressing with the aim for installation during LHC long shutdown 3 (2025--2027).
This contribution summarises the project motivation, its R\&D schedule, and will show selected highlights of recently accomplished project milestones, including full-scale engineering prototypes with dummy chips and small-scale, fully functional assemblies of functional, bent MAPS.
}

\keywords{Solid state detectors; Particle tracking detectors}

\begin{document}
\maketitle
\flushbottom

\section{ITS~3 project and R\&D activities}
\begin{figure}[tb]
\centerline{\includegraphics[width=.85\textwidth]{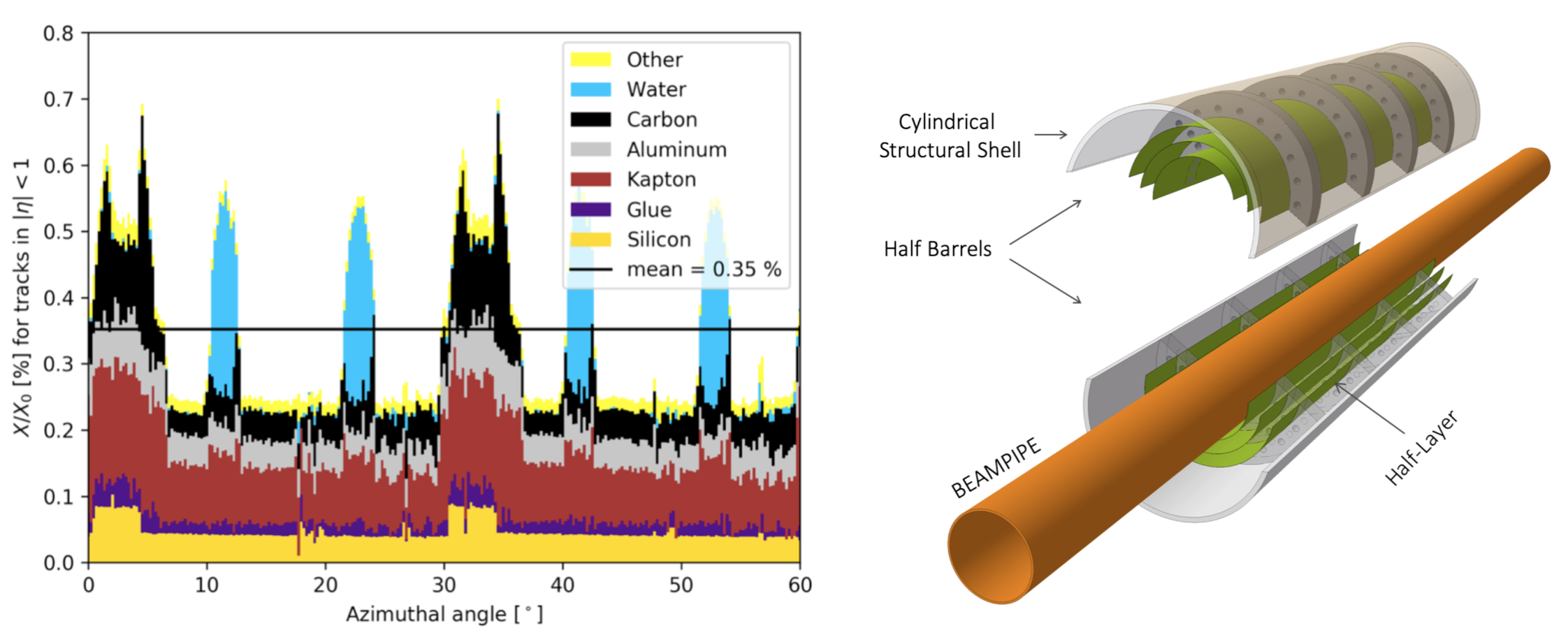}}
\caption{Left: Material budget angular distribution of two adjacent modules in the ITS~2 configuration. Right: proposed design for the inner barrel of the ITS in Run 4.}
\label{FigA}
\end{figure}

The ALICE Collaboration \cite{JINST} has just completed detector upgrades for data-taking during LHC Run~3 \cite{ALICEupLoI}. 
During this major enterprise the Inner Tracking System (ITS) used during LHC Run~1 and Run~2 has been replaced by a new detector, the ITS~2, based on Monolithic Active Pixel Sensors (MAPS) \cite{ITSUPTDR}. To achieve the pointing resolution needed for the defined ALICE physics program during LHC Run~3 and Run~4, the detector has been designed to have very high intrinsic spatial resolutions (5~$\mu$m), to be placed close to the interaction point (23~mm for the innermost layer), and to feature low material budget (0.35\%~X/X$_{0}$ for the inner layers). Details of the angular distribution of the material budget for two adjacent modules of the innermost layer of the ITS~2 can be seen in Figure~\ref{FigA} (left). The following considerations can be made: i) material budget is, to large extent, formed by passive components such as water cooling, carbon and kapton support structures, and aluminum wires; ii) a lot of irregularities, e.g. due to the overlap of adjacent modules (needed to grant hermeticity) or due to the presence of the water cooling pipes, are visible; iii) silicon contributes only about 15\% of the total material budget.
Based on these observations, ALICE has started a new R\&D programme towards next-generation inner-tracker layers, ITS~3, that aims to reduce the material budget to the minimum by removing the passive components and keeping just the silicon layer. Such a detector, designed to replace the three innermost layers of the ITS~2, would achieve an unprecedented low material budget of about 0.05\%~X/X$_{0}$ per layer \cite{ITS3LoI}. 

The construction of the detector would require (i) to reduce the power consumption below 20~mW/cm$^{2}$, in order to cool down the sensors only by airflow, (ii) to integrate power and data buses on the chip, in order to remove any other flex covering sensitive region, and (iii) to rely on the stiffness of large size, bent silicon wafers, in order to remove mechanical support structures. 
The first two points require the design of a new chip while the last point needs R\&D activities for sensor bending and light support development.
The composition of the new detector is shown in Figure~\ref{FigA} (right). Each layer is made up of two large dimension silicon chips (largest one having a surface area of 280~$\times$~94~mm$^2$) held in place using carbon foam structures and electrically connected only from a single side in a way that makes them compatible with the existing infrastructure of ITS~2. The installation of a new beam-pipe with smaller radius (inner radius 16~mm) and thickness (500~$\mu$m), will allow to place the layer even closer to the first interaction point (18~mm). 
Performance studies indicate an improvement by a factor 2 of the pointing resolution and of the stand-alone tracking efficiency (for $p_{\rm T} <$~100~MeV/c) with respect to ITS~2.
The improvement of the vertexing performance and the reduction of material budget will have a dramatic impact on the measurement of charm and beauty hadrons at low transverse momentum as well as on the measurement of low-mass and low-$p_{\rm T}$ dielectrons. 


\subsection{Detector integration}
\begin{figure}[tb]
\centerline{\includegraphics[width=.85\textwidth]{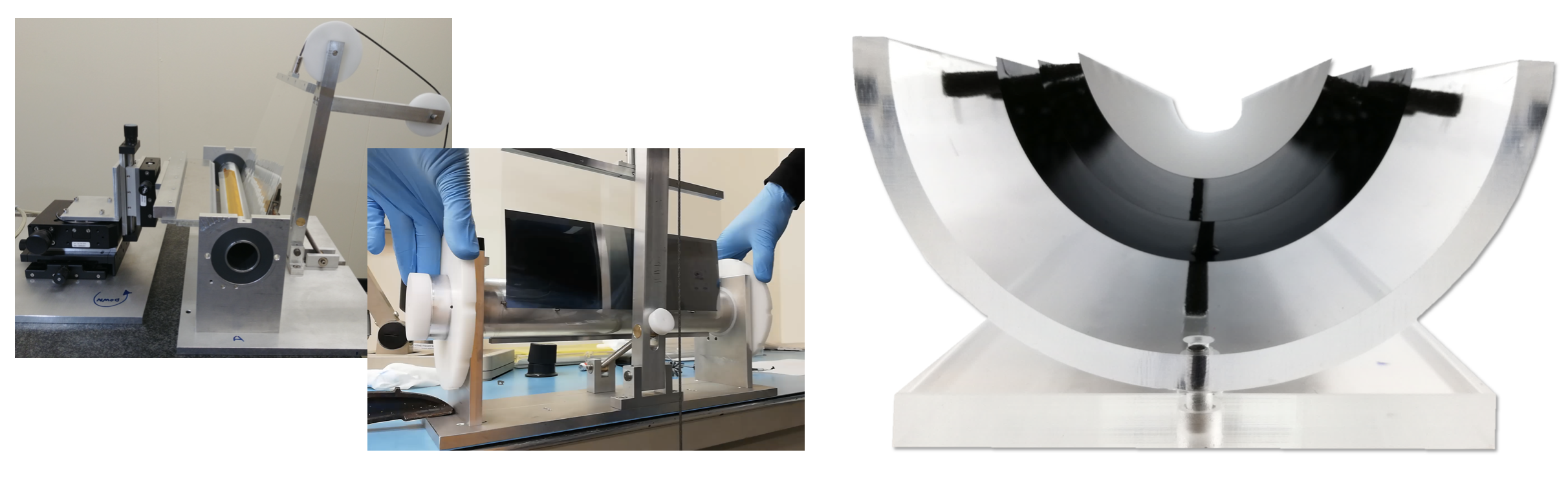}}
\caption{Left: Large dimension chips bending tool. Right: ITS~3 mechanical prototype assembly.}
\label{FigB}
\end{figure}

A 3-point bending test, made with ALPIDE \cite{ALPIDE} chips (MAPS sensors used for the assembly of the ITS~2) at four different thicknesses (between 100 and 30~$\mu$m), measuring the displacement from flat position as a function of the applied force, demonstrates that ITS~3 target radii are easily reachable with silicon thickness below 50~$\mu$m. 
As demonstrated with ALPIDE chips, the power consumption of the pixel matrix can stay quite low (40~mW/cm$^2$), concentrating most of it in the periphery where the readout electronics are integrated. A combination of air flow and contact dissipation with carbon foam support structures is a viable option to cool-down the sensors, removing the cooling pipes and coolant in the active volume. An extensive study established the best commercially available carbon foam materials to be used for mechanical supports, targeting an excellent thermal conductivity with a low density.
A full size mechanical prototype, using large dimension (50~$\mu$m thick) blank wafer, has been assembled, allowing the development of bending tools and verification of carbon foam support structures. In Figure~\ref{FigB} one can see how the chips are held in place using carbon foam wedges, whose material budget contribution is negligible. The assembly is now being characterised for its mechanical properties and shape imperfections, and serves as reference for further optimisation of the procedure.

\subsection{Bent sensor characterisation}
\begin{figure}[tb]
\centerline{\includegraphics[width=.85\textwidth]{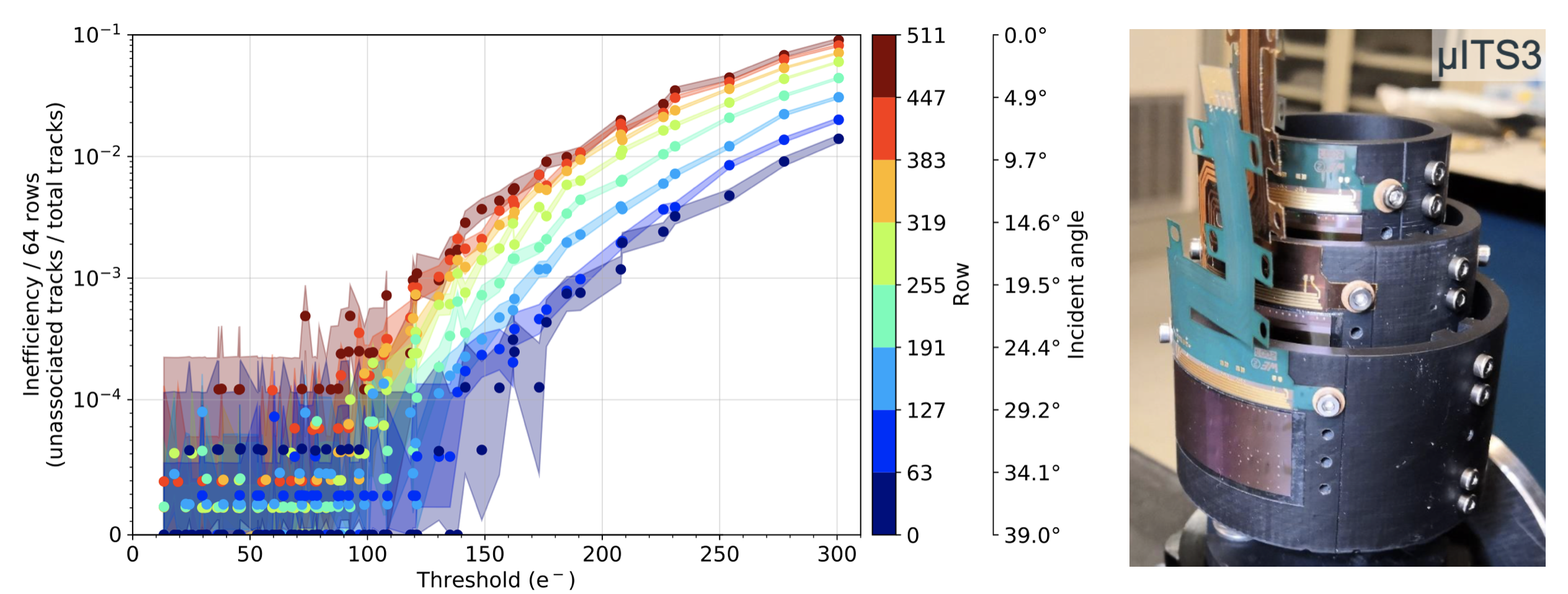}}
\caption{Left: hit inefficiency as a function of threshold for different rows and incidence angles as measured in a bent ALPIDE. Right: micro-ITS~3 assembly.}
\label{FigC}
\end{figure}

The ALPIDE sensor characterised in the laboratory showed no effect of bending on the performance; the noise level and number of dead pixels remained unchanged, and the difference in pixel threshold was negligible \cite{bentALPIDE}. The detection efficiency of a single bent ALPIDE sensor was measured at the DESY test beam facility, using a 5 GeV electron beam. In Figure~\ref{FigC} (left), it can be observed that below 100~e$^{-}$ the hit inefficiency is generally lower than 10$^{-4}$, independent of the incident angle or the position on the chip. An assembly, made of 6 ALPIDE sensors bent around cylinders of ITS~3 target radii, called $\mu$ITS3 (Figure \ref{FigC}, right), and with open windows in correspondence to the sensors, has been tested under beam to verify tracking and vertexing capabilities and the data analysis is currently ongoing.  

\subsection{Wafer-sensor design}
The required low value for the power consumption can be achieved by moving the sensor periphery, including the serial link, to the edge of the chip and by the usage of the 65~nm CMOS technology (ALPIDE was fabricated in 180~nm), allowing also implementation of smaller pixels. Tower Semiconductor is able to provide such technology on 300~mm-large wafers (180~nm is done on 200 mm wafers), allowing to produce sensors with the needed dimensions. Still this requires the application of a technique, known as stitching, that allows to overcome the chip mask or reticle size limitations (typically of the order of 2~$\times$~3~cm$^2$). New chip design in 65 nm reached its first milestone in June 2021 with the production of the multi-layer reticle 1 (MLR1) including first test structures like: transistor test structures, analog building blocks, various diode matrices and digital test matrices. Characterisation of these structures in laboratory and under beam is ongoing. Next important milestone will be the first engineer run including the first implementation of stitching technique.




%



\end{document}